\documentclass[preprint,prX]{revtex4}%
\usepackage{amsfonts}
\usepackage{amsmath}
\usepackage{amssymb}
\usepackage{graphicx}%
\begin{document}
\title[BCS approximation to the effective vertex]{BCS approximation to the effective vector vertex of superfluid fermions}
\author{L. B. Leinson}
\affiliation{Institute of Terrestrial Magnetism, Ionosphere and Radio Wave Propagation RAS,
142190 Troitsk, Moscow Region, Russia}

\begin{abstract}
We examine the effective interaction of nonrelativistic fermions with an
external vector field in superfluid systems. In contrast to the complicated
vertex equation, usually used in this case, we apply the approach which does
not employ an explicit form of the pairing interaction. This allows to obtain
a simple analytic expression for the vertex function only in terms of the
order parameter and other macroscopic parameters of the system.

We use this effective vertex to analyze the linear response function of the
superfluid medium at finite temperatures. At the time-like momentum transfer,
the imaginary part of the response function is found to be proportional to
$V_{F}^{4}$, i.e. the energy losses through vector currents are strongly
suppressed. As an application, we calculate the neutrino energy losses through
neutral weak currents caused by the pair recombination in the superfluid
neutron matter at temperatures lower than the critical one for the $^{1}S_{0}$
pairing. This approach confirms a strong suppression of the neutrino energy
losses as predicted in Ref. \cite{LP06}.

\end{abstract}
\startpage{1}
\maketitle





\section{Introduction}

Interactions among fermions in a superdense medium substantially modify their
coupling to external fields. This problem is of a great importance both for a
laboratory superconductors and for a superfluid baryon matter of neutron
stars, where the thermal breaking and recombination of Cooper pairs is
considered as a possible dominant mechanism of the neutrino energy losses
through neutral weak currents. The rate of reaction, in the $^{1}S_{0}$
superfluid neutron matter, was for the first time estimated in Ref.
\cite{FRS76}. Recent analysis \cite{LP06} has shown however that the approach
used in this calculation contradicts the hypothesis of conservation of the
vector current in weak interactions and therefore considerably overestimates
the neutrino energy losses.

An alternative calculation of the same process was suggested in Ref.
\cite{Vosk}, where the Green function technique was used to connect the rate
of neutrino emission with the current-current correlation function, also
called the retarded weak polarization tensor, and the corrections due to the
strong interactions have been incorporated in the weak vertex function. The
estimate was based on the idea that the dominant response of nonrelativistic
baryons is due to the coupling of the baryon density to neutrinos. In this
case the relevant input for the calculation is the imaginary part of the
temporal component of the retarded vector-vector polarization tensor,
Im$\Pi^{00}\left(  \omega,\mathbf{q}\right)  ,$which was estimated by the
authors in the kinematical domain $2\Delta<\omega<\infty$ and $\mathbf{q}=0$,
to obtain Im$\Pi^{00}\left(  \omega>2\Delta,\mathbf{0}\right)  \neq0$, where
$\Delta$ is the superfluid energy gap, and $q=\left(  \omega,\mathbf{q}%
\right)  $ is the transferred four momentum. It is easy to see, that this
estimate contradicts the current continuity condition, $\omega\Pi^{00}\left(
\omega,\mathbf{q}\right)  =q_{i}\Pi^{i0}\left(  \omega,\mathbf{q}\right)  $,
which tells us $\Pi^{00}\left(  \omega,\mathbf{0}\right)  =0$ when
$\omega>2\Delta$.

A one more calculation was recently done in Ref. \cite{Sedrakian} where the
vertex renormalization is reduced to a study of the effective three-point
vertices that sum-up particle-hole irreducible ladders in the scalar channel.
The vector channel of the particle-hole interactions was parametrized by a
constant (the lowest order Landau parameter) and the ideal summation is
performed in the random phase approximation. The result of this calculation is
also incorrect. Indeed, the whole longitudinal polarization function, as given
by Eq. (35) and/or by Eq. (48) of this work, vanishes when the gap goes to
zero. Thus, at temperatures higher the critical one for neutron pairing, both
the real and imaginary part of the polarization function vanish instead of
going over to those of normal (unpaired) Fermi liquid

A reasonable calculation of the medium response to external vector fields must
satisfy the current continuity conditions. Neglect of this requirement leads
to a considerable error in the estimate of the longitudinal response function.
It is well known that the current conservation in superfluid fermionic systems
is fulfilled in a nontrivial way. The interest to this problem has arisen many
years ago in connection with the gauge invariance of the
Bardeen-Cooper-Schrieffer (BCS) theory \cite{Bogoliubov}, \cite{Anderson},
\cite{Nambu}, \cite{Amb}, \cite{Migdal}. But the problem is topical up to now
\cite{LP06}, \cite{Arseev}, \cite{Reddy}.

It was realized that the current continuity equation can be satisfied if the
interaction among quasiparticles is incorporated in the coupling vertex to the
same degree of approximation as the self-energy effect is included in the
quasiparticle. This prescription yields the vertex equation explicitly
incorporating two physical inputs -- the pairing interaction and the energy
gap which are not independent but connected by the gap equation. Because the
realistic pairing interactions are momentum dependent \cite{Tamagaki},
\cite{Dean}, \cite{Clark}, particularly due to a short-range repulsive core,
the above set of equations admits only numerical solution.

On the other hand it is well known that the BCS theory of superfluidity needs
the particular form of the pairing interaction only for a calculation of the
order parameter (the energy gap $\Delta$). The latter completely defines all
the properties of a superconductor. Thus, it is reasonable to expect that, in
the BCS approximation, the vertex function can be expressed only in terms of
the order parameter, and other macroscopic parameters of the superfluid
system, without of using of an explicit form of the pairing interaction. Up to
now, the relevant solution of the vertex equation is obtained \cite{Nambu},
\cite{Amb} only for the case when both the transferred energy and momentum are
smaller than the superfluid energy gap, $\omega,\left\vert \mathbf{q}%
\right\vert <\Delta$, and only for zero temperature, $T=0$. In applications,
however, we frequently deal with the time-like momentum transfer at finite
temperature $T>0$.

In this paper we focus on the analytic calculation of the corresponding
effective vertex valid at finite temperatures and for $\omega$ and
$\mathbf{q}$, satisfying only $\omega\ll\mu,\left\vert \mathbf{q}\right\vert
\ll\mathsf{p}_{F}$, where $\mu=\mathsf{p}_{F}^{2}/2M^{\ast}$ and
$\mathsf{p}_{F}$ are the effective chemical potential and Fermi momentum of
interacting fermions, respectively. Our calculation explicitly ensures the
current conservation, using method that imposes this from the start. The
essential idea of the method is as follows. We know that the external
longitudinal field modifies the order parameter in the system. The
corresponding self-consistent correction can be found directly from the
condition of the current conservation. As was repeatedly discussed in the
theory of superconductors, this procedure is equivalent to the above summation
of the vertex corrections \cite{Amb}, \cite{Arseev}.

The vector vertex in the BCS approximation is to be considered as a starting
point for incorporation of residual interactions in order to provide the
current conservation in the superfluid system. In turn, the residual
interactions, evaluated in the random phase approximation are known to take
into account the effects of the medium polarization (see e.g. \cite{Clark},
\cite{L01}), which renormalizes the sound velocity and the normal part of the
vector vertex. This is beyond the scope of our consideration.

The paper is organized as follows. Section 2 outlines some well known
properties of normal nonrelativistic Fermi liquids at low temperatures and
introduces the notations used in the following. In section 3, we derive the
general expression for the effective vector vertex of electrically neutral
fermions in the pair-correlated system and evaluate the obtained expression in
the quasiparticle approximation. In section 4, we consider the linear response
of the superfluid medium in the vector channel. As an application, in section
5, we evaluate the neutrino energy losses through neutral weak currents caused
by the pair recombination of neutrons in the crust of neutron stars. Section 6
contains a short summary of our findings and the conclusion.

We use the system of units $\hbar=c=1$, and the Boltzmann constant $k_{B}=1$.

\section{Normal Fermi liquid}

In order to introduce the approach used in the following consideration this
section outlines some well known properties of normal nonrelativistic Fermi
liquids at low temperatures. The degenerate fermion systems are commonly
treated in the quasiparticle approximation. Within this framework, near the
Fermi surface, the inverse Green function of normal (unpaired) fermions takes
the quasiparticle form
\begin{equation}
G_{\mathrm{N}}^{-1}\left(  \varepsilon,\mathbf{p}\right)  \simeq
\varepsilon-\xi_{\mathbf{p}}, \label{Gqp}%
\end{equation}
where
\begin{equation}
\xi_{\mathbf{p}}=\frac{\mathsf{p}^{2}}{2M^{\ast}}-\frac{\mathsf{p}_{F}^{2}%
}{2M^{\ast}}\simeq\frac{\mathsf{p}_{F}}{M^{\ast}}(\mathsf{p}-\mathsf{p}_{F}),
\label{ksi}%
\end{equation}
is the energy of a quasiparticle relative to the Fermi level. The effective
mass $M^{\ast}$ of a quasiparticle is connected to the Fermi momentum
$\mathsf{p}_{F}$ by means of the Fermi velocity $V_{F}=$ $\mathsf{p}%
_{F}/M^{\ast}$ which is small in a nonrelativistic fermion system.

In the lowest-order expansion (\ref{Gqp}) the wave-function renormalization
which accounts for the next-to-leading term in the expansion around the
Fermi-energy is set to unity. For the same reason we neglect the imaginary
part of the quasiparticle self-energy $\sim\left\vert \varepsilon\right\vert
\varepsilon$.

It is convenient to work in the particle-hole picture (known as the
Nambu-Gor'kov formalism) by introducing the quasiparticle fields as
two-component (particle-hole) objects $\Psi_{p}$ and using the Pauli matrices
$\hat{\tau}_{i}$ ($i=1,2,3$)\ operating in the Nambu-Gor'kov space (see e.g.
\cite{Nambu}). Then the Hamiltonian of the system of free quasi-particles
takes the form
\begin{equation}
H_{0}=\sum_{p}\Psi_{p}^{\dagger}\xi_{\mathbf{p}}\hat{\tau}_{3}\Psi_{p}
\label{Ham}%
\end{equation}
which corresponds to excited states in the particle-hole picture, while the
ground state (vacuum) is the state where all negative energy "quasi-particles"
($\epsilon<0$) are occupied and no positive energy particles exist.

In terms of these fields, the vector vertex of a quasiparticle becomes the
diagonal matrix:
\begin{equation}
\hat{\gamma}_{\mu}(p+q\mathbf{,}p)=\left(
\begin{array}
[c]{cc}%
\gamma_{\mu}\left(  p+q,p\right)  & 0\\
0 & -\gamma_{\mu}^{\dagger}\left(  p+q,p\right)
\end{array}
\right)  , \label{gamb}%
\end{equation}
where the quasiparticle component is given by
\[
\gamma_{\mu}\left(  p+q,p\right)  =\left(  1,\frac{1}{M^{\ast}}\left(
\mathbf{p+}\frac{1}{2}\mathbf{q}\right)  \right)
\]
and the hole component is $\gamma_{\mu}^{\dagger}\left(  p+q,p\right)
=\gamma_{\mu}\left(  -p,-p-q\right)  $.

Using the Pauli matrices this can be recast as
\begin{equation}
\hat{\gamma}^{\mu}=\left\{
\begin{array}
[c]{cc}%
\hat{\tau}_{3} & \mathrm{if\ }\mu=0,\ \ \ \ \ \ \ \ \ \ \ \ \\
\frac{1}{M^{\ast}}\left(  \mathbf{p+}\frac{1}{2}\mathbf{q}\right)  &
\mathrm{if\ }\mu=i=1,2,3
\end{array}
\right.  . \label{gam}%
\end{equation}
The inverse quasiparticle propagator takes the following matrix form
\begin{equation}
\hat{G}_{0}^{-1}\left(  \varepsilon,\mathbf{p}\right)  =\varepsilon
-\xi_{\mathbf{p}}\hat{\tau}_{3}. \label{Gqpm}%
\end{equation}

\section{Superfluidity effects}

In the particle-hole picture, the pairing represents a quasiparticle
transition into a hole (and a correlated pair). Therefore the self-energy
arising due to the pairing interaction has the off-diagonal form in the
Nambu-Gor'kov space. In the absence of external fields this anomalous
self-energy is real-valued and can be written in terms of the energy gap
$\Delta$ arising in the quasiparticle spectrum. Then the BCS inverse
propagator of a quasiparticle has the following form (see e.g. \cite{Migdal}):%
\begin{equation}
\hat{G}^{-1}\left(  \varepsilon,\mathbf{p}\right)  =\varepsilon-\xi
_{\mathbf{p}}\hat{\tau}_{3}-\Delta\hat{\tau}_{1}, \label{iG}%
\end{equation}

The the effective vertex in the pair-correlated system can be written as%
\begin{equation}
\hat{\Gamma}_{\mu}=\hat{\gamma}_{\mu}+\frac{\partial\hat{\Sigma}^{\left(
1\right)  }}{\delta V^{\mu}}. \label{Gdef}%
\end{equation}
where $\hat{\gamma}_{\mu}$ is the "normal" vertex, as given by Eq.
(\ref{gamb}), and $\hat{\Sigma}^{\left(  1\right)  }$ is the linear correction
to the anomalous self-energy of a quasiparticle in the external vector field
$V^{\mu}$. Apparently this correction is due to varying of the energy gap in
the external field. By taking the new gap in the form
\[
\tilde{\Delta}=\Delta+\Delta^{\left(  1\right)  }\left(  V^{\mu}\right)
\]
we can write
\[
\hat{\Sigma}^{\left(  1\right)  }=\left(
\begin{array}
[c]{cc}%
0 & \Delta^{\left(  1\right)  }\left(  V^{\mu}\right) \\
\Delta^{\ast\left(  1\right)  }\left(  V^{\mu}\right)  & 0
\end{array}
\right)  .
\]

Now we use the fact that, in the case $\mathsf{q}\ll\mathsf{p}_{F}$, the
linear effect of the weak field perturbation is to change the phase
$\Phi\left(  V^{\mu}\right)  $ but not the magnitude of the gap parameter
\cite{Amb}, \cite{Arseev}. The field-induced correction to the amplitude of
the gap is proportional to $\Delta/\mu\ll1$ and thus can be neglected. We
therefore write the energy gap as $\tilde{\Delta}=\Delta\exp i\Phi\left(
V^{\mu}\right)  $, where $\Delta$ is the (real) gap in the absence of external
field. Apparently the linear correction to the gap $\Delta^{\left(  1\right)
}=\tilde{\Delta}-\Delta$ is proportional to the external field and can be
written as
\begin{align}
\Delta^{\left(  1\right)  }\left(  V^{\mu}\right)   &  \equiv\Delta
\cdot\left(  e^{i\Phi\left(  V^{\mu}\right)  }-1\right)  \simeq i\Delta
\cdot\Phi\left(  V^{\mu}\right)  \simeq-Q_{\mu}\left(  q\right)  V^{\mu
}\left(  q\right)  ,\nonumber\\
\Delta^{\ast\left(  1\right)  }\left(  V^{\mu}\right)   &  \equiv\Delta
\cdot\left(  e^{-i\Phi\left(  V^{\mu}\right)  }-1\right)  \simeq-i\Delta
\cdot\Phi\left(  V^{\mu}\right)  \simeq Q_{\mu}\left(  q\right)  V^{\mu
}\left(  q\right)  , \label{Del1}%
\end{align}
where $Q_{\mu}\left(  q\right)  $ is the unknown kernel connecting the
correction to the energy gap with the external weak field. Then the effective
vertex becomes%
\begin{equation}
\hat{\Gamma}_{\mu}(p+q\mathbf{;}p)=\left(
\begin{array}
[c]{cc}%
\gamma_{\mu}(p+q\mathbf{,}p) & -Q_{\mu}(q)\\
Q_{\mu}(q) & -\gamma_{\mu}^{\dagger}\left(  p+q,p\right)
\end{array}
\right)  , \label{GAM}%
\end{equation}
and the problem reduces to calculation of the unknown vector function $Q_{\mu
}\left(  \omega,\mathbf{q}\right)  .$

The relation between the modified vertex $\Gamma^{\mu}$ and the quasiparticle
propagator $\hat{G}$ is given by the matrix Ward identity \cite{Schr}:
\begin{equation}
q^{\mu}\hat{\Gamma}_{\mu}\left(  p+q,p\right)  =\hat{G}^{-1}\left(
p+q\right)  \hat{\tau}_{3}-\hat{\tau}_{3}\hat{G}^{-1}\left(  p\right)  .
\label{Ward}%
\end{equation}
Applying this identity to expressions (\ref{GAM}) and (\ref{iG}) we obtain the
following well known relation \cite{Migdal}:%
\begin{equation}
q_{\nu}Q^{\nu}\left(  q\right)  =-2\Delta. \label{WR}%
\end{equation}

For further progress, let us consider the medium response $\Pi^{\mu\nu}(q)$ to
the external vector field. This calculation needs the quasiparticle propagator
which, in contrast to the inverse propagator (\ref{iG}), must be uniquely
defined at finite temperature. Since the matter is assumed in thermal
equilibrium at some temperature, we employ the Matsubara calculation
technique. In this case the quasiparticle propagator can be obtained by
inverting Eq. (\ref{iG}), and substituting $\varepsilon=i\pi\left(
2n+1\right)  T$ with $n=0,\pm1,\pm2....$ Then the retarded polarization tensor
$\Pi^{\mu\nu}(q)$ is given by the analytical continuation of the following
Matsubara sums (see e.g. \cite{Schr})
\begin{align}
\Pi^{\mu\nu}(\omega_{m},\mathbf{q})  &  =T\sum_{p_{n}}\int\frac{d^{3}p}%
{(2\pi)^{3}}\mathrm{Tr}\left[  \hat{\gamma}^{\mu}\left(  p_{n}+\omega
_{m},\mathbf{p+q;}p_{n},\mathbf{p}\right)  \hat{G}\left(  p_{n}+\omega
_{m},\mathbf{p+q}\right)  \right. \nonumber\\
&  \times\left.  \hat{\Gamma}^{\nu}\left(  p_{n}+\omega_{m},\mathbf{p+q}%
;p_{n},\mathbf{p}\right)  \hat{G}\left(  p_{n},\mathbf{p}\right)  \right]
\label{PIM}%
\end{align}
onto the upper-half plane of the complex variable $\omega$. Here $p_{n}%
=\pi\left(  2n+1\right)  T$, and\ $\omega_{m}=2\pi mT$, with $m,n=0,\pm
1,\pm2...$, are the fermionic and bosonic Matsubara frequency, respectively;
and the quasiparticle propagator has the following form
\begin{equation}
\hat{G}\left(  p_{n},\mathbf{p}\right)  \equiv\left(
\begin{array}
[c]{cc}%
G\left(  p_{n},\mathbf{p}\right)  & F^{\dagger}\left(  p_{n},\mathbf{p}\right)
\\
F\left(  p_{n},\mathbf{p}\right)  & -G^{\dagger}\left(  p_{n},\mathbf{p}%
\right)
\end{array}
\right)  . \label{G}%
\end{equation}
The particle and hole components are given by the on-diagonal elements, with
$G^{\dagger}\left(  p_{n},\mathbf{p}\right)  =G\left(  -p_{n},-\mathbf{p}%
\right)  $ while the off-diagonal elements, $F^{\dagger}\left(  p_{n}%
,\mathbf{p}\right)  =F\left(  -p_{n},-\mathbf{p}\right)  $, represent the
anomalous contribution caused by the pairing interaction. According to Eq.
(\ref{iG}), in the BCS approximation the components are given by the following
expressions \cite{AGD}:%
\begin{align}
G\left(  p_{n},\mathbf{p}\right)   &  =\frac{-ip_{n}-\xi_{\mathbf{p}}}%
{p_{n}^{2}+\varepsilon_{\mathbf{p}}^{2}},\ \ \ \ F\left(  p_{n},\mathbf{p}%
\right)  =\frac{\Delta}{p_{n}^{2}+\varepsilon_{\mathbf{p}}^{2}},\nonumber\\
F^{\dagger}\left(  p_{n},\mathbf{p}\right)   &  =F\left(  p_{n},\mathbf{p}%
\right)  ,\ \ \ \ \ \ G^{\dagger}\left(  p_{n},\mathbf{p}\right)
=\frac{ip_{n}-\xi_{\mathbf{p}}}{p_{n}^{2}+\varepsilon_{\mathbf{p}}^{2}},
\label{GF}%
\end{align}
where $\varepsilon_{\mathbf{p}}$ is the energy of a quasiparticle
\begin{equation}
\varepsilon_{\mathbf{p}}=\sqrt{\xi_{\mathbf{p}}^{2}+\Delta^{2}}. \label{eps}%
\end{equation}
It is important to notice that the magnitude of the energy gap in the above
expressions depends on the temperature, $\Delta=\Delta\left(  T\right)  $. We
assume that this function is found from the corresponding gap equation.

Using expressions (\ref{G}), (\ref{GAM}) in Eq. (\ref{PIM}) gives
\begin{equation}
\Pi_{\mu\nu}\left(  q\right)  =\Lambda_{\mu\nu}\left(  q\right)  +\Lambda
_{\mu}\left(  q\right)  Q_{\nu}\left(  q\right)  , \label{Pi}%
\end{equation}
where the one-loop integrals $\Lambda_{\mu\nu}\left(  \omega,\mathbf{q}%
\right)  $ and $\Lambda_{\mu}\left(  \omega,\mathbf{q}\right)  $ are defined
as the analytical continuation of the following Matsubara sums:
\begin{align}
\Lambda_{\mu\nu}\left(  \omega_{m},\mathbf{q}\right)   &  =T\sum_{p_{n}}%
\int\frac{d^{3}p^{\prime}}{\left(  2\pi\right)  ^{3}}\left[  \gamma_{\mu
}\gamma_{\nu}G\left(  p_{n}+\omega_{m},\mathbf{p}^{\prime}\mathbf{+q}\right)
G\left(  p_{n},\mathbf{p}^{\prime}\right)  \right. \nonumber\\
&  +\gamma_{\mu}^{\dagger}\gamma_{\nu}^{\dagger}G^{\dagger}\left(
p_{n}+\omega_{m},\mathbf{p}^{\prime}\mathbf{+q}\right)  G^{\dagger}\left(
p_{n},\mathbf{p}^{\prime}\right) \nonumber\\
&  \left.  -\left(  \gamma_{\mu}\gamma_{\nu}^{\dagger}+\gamma_{\mu}^{\dagger
}\gamma_{\nu}\right)  F\left(  p_{n},\mathbf{p}^{\prime}\right)  F\left(
p_{n}+\omega_{m},\mathbf{p}^{\prime}\mathbf{+q}\right)  \right]  \label{MN}%
\end{align}%
\begin{align}
\Lambda_{\mu}\left(  \omega_{m},\mathbf{q}\right)   &  =T\sum_{p_{n}}\int
\frac{d^{3}p^{\prime}}{\left(  2\pi\right)  ^{3}}\left[  \gamma_{\mu}G\left(
p_{n}+\omega_{m},\mathbf{p}^{\prime}\mathbf{+q}\right)  F\left(
p_{n},\mathbf{p}^{\prime}\right)  \right. \nonumber\\
&  +\gamma_{\mu}^{\dagger}F\left(  p_{n}+\omega_{m},\mathbf{p}^{\prime
}\mathbf{+q}\right)  G^{\dagger}\left(  p_{n},\mathbf{p}^{\prime}\right)
\nonumber\\
&  -\gamma_{\mu}^{\dagger}G^{\dagger}\left(  p_{n}+\omega_{m},\mathbf{p}%
^{\prime}\mathbf{+q}\right)  F\left(  p_{n},\mathbf{p}^{\prime}\right)
\nonumber\\
&  \left.  -\gamma_{\mu}F\left(  p_{n}+\omega_{m},\mathbf{p}^{\prime
}\mathbf{+q}\right)  G\left(  p_{n},\mathbf{p}^{\prime}\right)  \right]  .
\label{M}%
\end{align}

The unknown vector function $Q_{\nu}\left(  q\right)  $ can be found from the
requirement that the polarization tensor (\ref{Pi}) obeys the current
conservation, $\Pi^{\mu\nu}q_{\nu}=0$, and $q_{\nu}\Pi^{\nu\mu}=0$. The first
of these relations is satisfied automatically if the effective vertex
$\hat{\Gamma}_{\mu}$ obeys the Ward identity (\ref{Ward}), the second one is
to be satisfied by a proper choice of the function $Q^{\mu}\left(  q\right)  $.

With the aid of relation (\ref{WR}) the above conditions can be written as the
two coupled equations%
\begin{equation}
\Lambda^{\mu\nu}\left(  q\right)  q_{\nu}-2\Delta\ \Lambda^{\mu}\left(
q\right)  =0, \label{1}%
\end{equation}%
\begin{equation}
q_{\nu}\Lambda^{\nu\mu}\left(  q\right)  +q_{\nu}\Lambda^{\nu}\left(
q\right)  Q^{\mu}\left(  q\right)  =0. \label{qLmn}%
\end{equation}
Using the symmetry property, $q_{\nu}\Lambda^{\nu\mu}=\Lambda^{\mu\nu}q_{\nu}%
$, we find%
\begin{equation}
Q^{\mu}\left(  q\right)  \ =-\frac{2\Delta}{q_{\lambda}\Lambda^{\lambda
}\left(  q\right)  }\ \Lambda^{\mu}\left(  q\right)  . \label{Qmu}%
\end{equation}
Inserting this in Eq. (\ref{GAM}) we obtain the effective vector vertex in the
following form:
\begin{equation}
\hat{\Gamma}_{\mu}(p+q\mathbf{;}p)=\hat{\gamma}_{\mu}(p+q\mathbf{,}%
p)-\frac{2i\Delta\hat{\tau}_{2}}{q_{\lambda}\Lambda^{\lambda}\left(  q\right)
}\ \Lambda_{\mu}\left(  q\right)  . \label{GAMef}%
\end{equation}

Thus the BCS approximation reduces to the calculation of the vector function
$\Lambda^{\mu}\left(  q\right)  $. In Eqs. (\ref{MN}), (\ref{M}), the
summation over the Matsubara fermionic frequency can be performed exactly. The
subsequent analytical continuation yields
\begin{align}
~\Lambda_{00}\left(  \omega,\mathbf{q}\right)   &  =-\int\frac{d^{3}%
\mathsf{p}}{2\left(  2\pi\right)  ^{3}}\left[  \left(  1-\frac{\xi
_{\mathsf{p}}\xi_{\mathbf{p+q}}-\Delta^{2}}{\varepsilon_{\mathsf{p}%
}\varepsilon_{\mathbf{p+q}}}\right)  \Phi_{+}\right. \nonumber\\
&  \left.  -~\left(  1+\frac{\xi_{\mathsf{p}}\xi_{\mathbf{p+q}}-\Delta^{2}%
}{\varepsilon_{\mathsf{p}}\varepsilon_{\mathbf{p+q}}}\right)  \Phi_{-}\right]
, \label{L00}%
\end{align}%
\begin{equation}
\Lambda_{0}\left(  \omega,\mathbf{q}\right)  =-\int\frac{d^{3}\mathsf{p}%
}{2\left(  2\pi\right)  ^{3}}\frac{\omega\Delta}{\varepsilon_{\mathsf{p}%
}\varepsilon_{\mathbf{p+q}}}\left(  \Phi_{+}+\Phi_{-}\right)  , \label{L0}%
\end{equation}%
\begin{equation}
\Lambda_{i}\left(  \omega,\mathbf{q}\right)  =-\frac{1}{M^{\ast}}\int
\frac{d^{3}\mathsf{p}}{2\left(  2\pi\right)  ^{3}}\left(  \mathsf{p}%
_{i}\mathbf{+}\frac{1}{2}\mathsf{q}_{i}\right)  \frac{\Delta\left(
\xi_{\mathbf{p+q}}-\xi_{\mathsf{p}}\right)  }{\varepsilon_{\mathsf{p}%
}\varepsilon_{\mathbf{p+q}}}\left(  \Phi_{+}+\Phi_{-}\right)  . \label{Lami}%
\end{equation}%
\begin{equation}
q_{i}\Lambda_{i}\left(  \omega,\mathbf{q}\right)  =-\int\frac{d^{3}\mathsf{p}%
}{2\left(  2\pi\right)  ^{3}}\frac{\Delta\left(  \xi_{\mathbf{p+q}}%
-\xi_{\mathsf{p}}\right)  ^{2}}{\varepsilon_{\mathsf{p}}\varepsilon
_{\mathbf{p+q}}}\left(  \Phi_{+}+\Phi_{-}\right)  . \label{Li}%
\end{equation}
where the following notations are introduced:%
\begin{equation}
\Phi_{\pm}=\frac{\varepsilon_{\mathbf{p+q}}\pm\varepsilon_{p}}{\left(
\varepsilon_{\mathbf{p+q}}\pm\varepsilon_{p}\right)  ^{2}-\left(
\omega+i0\right)  ^{2}}\left(  \tanh\frac{\varepsilon_{p}}{2T}\pm\tanh
\frac{\varepsilon_{\mathbf{p+q}}}{2T}\right)  . \label{poles}%
\end{equation}

Eq. (\ref{GAMef}) is the general result valid for arbitrary momentum
transfers. By the direct substitution one can verify that this effective
vertex satisfies the Ward identity (\ref{Ward}), providing the current
conservation in the system.

As we can see, in the BCS approximation, only the longitudinal components of
the vertex are modified because the transverse components of $\Lambda_{i}$
vanish at the angle integration. Making use of this fact we need to calculate
only the temporal component of the vertex because the remaining components can
be found from the current continuity condition.

General expression for the vertex function, as given by Eqs. (\ref{GAMef}) -
(\ref{poles}), has a complicated form. Significant simplification is possible,
however, due to the fact that the quasiparticles are nonrelativistic, i.e
$V_{F}\ll1$ and because we are interested in $\mathsf{q}\ll\mathsf{p}_{F}$.
The latter condition implies $\mathsf{q}/M^{\ast}=V_{F}\mathsf{q/p}_{F}\ll
V_{F}$ and we may neglect the quasiparticle recoil. Indeed, in the cases
typical for astrophysical applications one has $V_{F}\sim0.1$ while
$\mathsf{q}/M^{\ast}\sim T/M^{\ast}\sim10^{-4}\div10^{-3}$. Under these
conditions the contributions, caused by the recoil are estimated as
$\mathsf{q}^{2}/\left(  M^{\ast}\omega\right)  \sim\left(  \mathsf{q}/M^{\ast
}\right)  \left(  q/\omega\right)  \sim V_{F}\left(  \mathsf{q}/\mathsf{p}%
_{F}\right)  \left(  q/\omega\right)  $.

We have
\begin{equation}
\xi_{\mathbf{p+q}}-\xi_{\mathsf{p}}=\mathsf{q}V_{F}\cos\theta+\frac
{\mathsf{q}^{2}}{2M^{\ast}}=\mathsf{q}V_{F}\left(  \cos\theta+\frac
{\mathsf{q}}{2\mathsf{p}_{F}}\right)  \label{b}%
\end{equation}
where $\theta$ is the angle between $\mathbf{p}$ and $\mathbf{q}$ momenta.
Insertion in Eq. (\ref{Li}) gives%
\[
q_{i}\Lambda_{i}\left(  \omega,\mathbf{q}\right)  =-\mathsf{q}^{2}V_{F}%
^{2}\int\left(  \cos\theta+\frac{\mathsf{q}}{2\mathsf{p}_{F}}\right)
^{2}\frac{d^{3}\mathsf{p}}{\left(  2\pi\right)  ^{3}}\frac{\Delta
}{2\varepsilon_{\mathsf{p}}\varepsilon_{\mathbf{p+q}}}\left(  \Phi_{+}%
+\Phi_{-}\right)  .
\]
In this expression, contributions due to the odd powers of cosine vanish after
angle integrations. To the lowest order in small parameters we may take the
integrand at $\mathbf{q}=0$ to find the following relation%
\[
q_{i}\Lambda_{i}\left(  \omega,\mathbf{q}\right)  \simeq\frac{1}{\omega
}\mathsf{q}^{2}\frac{V_{F}^{2}}{3}\Lambda_{0}\left(  \omega,\mathbf{0}\right)
\]
valid up to accuracy $V_{F}^{3}$. Apparently to the same accuracy we can
write
\begin{equation}
q_{i}\Lambda_{i}\left(  \omega,\mathbf{q}\right)  \simeq\frac{1}{\omega
}\mathsf{q}^{2}c_{s}^{2}\Lambda_{0}\left(  \omega,\mathbf{q}\right)  ,
\label{qiLi}%
\end{equation}
where $c_{s}\equiv V_{F}/\sqrt{3}$ is the sound velocity in the Fermi gas.

Insertion in Eq. (\ref{GAMef}) immediately gives%
\begin{equation}
\hat{\Gamma}_{0}(p+q\mathbf{;}p)\simeq\hat{\tau}_{3}-2i\Delta\hat{\tau}%
_{2}\frac{\omega}{\omega^{2}-\mathsf{q}^{2}c_{s}^{2}}\ , \label{GAM0}%
\end{equation}
The space component of the longitudinal vertex can be \ found from the Ward
identity:
\begin{equation}
\mathbf{q\hat{\Gamma}}\simeq\xi_{\mathbf{p+q}}-\xi_{\mathbf{p}}-2i\Delta
\hat{\tau}_{2}\frac{q^{2}c_{s}^{2}}{\omega^{2}-\mathsf{q}^{2}c_{s}^{2}}.
\label{GAMq}%
\end{equation}
It is necessary to stress that the above expressions are obtained from the
exact expression Eq. (\ref{GAMef}) to accuracy $V_{F}^{2}\ll1$ valid in the
nonrelativistic system. To this accuracy the temperature dependence of the
effective vector vertex is restricted to the gap function $\Delta\left(
T\right)  $.

It is interesting to notice that the approximate Eqs. (\ref{GAM0}) and
(\ref{GAMq}) reproduce formally the vertex functions as obtained in Ref.
\cite{Nambu} for the case of small transferred energy and momentum,
$\omega,\mathsf{q}\ll\Delta\mathsf{,}$ and zero temperature. In contrast, our
analysis shows that Eqs. (\ref{GAM0}), (\ref{GAMq}) are valid under
considerably weaker conditions $\omega\ll\mu\equiv\mathsf{p}_{F}^{2}/2M^{\ast
}$, $\mathsf{q}\ll\mathsf{p}_{F}$ and for a time-like momentum transfer,
$\left\vert \mathbf{q}\right\vert <\omega$, as well. The energy gap
$\Delta=\Delta\left(  T\right)  $, in Eqs. (\ref{GAM0}), and (\ref{GAMq}),
depends on the temperature below the critical temperature $T_{c}$. At
temperatures $T>T_{c}$ the gap vanishes. In this case, Eqs. (\ref{GAM0}), and
(\ref{GAMq}) give the vector vertex (\ref{gam}), as obtained for normal Fermi liquids.

As is well known, the poles in the vertex function at $\omega^{2}%
=\mathsf{q}^{2}c_{s}^{2}$ indicate the existence of the collective
acoustic-like excitations in the condensate propagating with a velocity
$c_{s}$. The second term in (\ref{GAM0}), and (\ref{GAMq}) is the result of
the coupling of $\hat{\tau}_{3}$ to the collective mode. This can be
understood in the following way. $\hat{\Gamma}_{\mu}$ contains matrix elements
for creation or annihilation of a pair out of the vacuum. This process can go
through the virtual collective intermediate state. This collective
contribution plays the important role in the conservation of the vector
current in superfluid systems.

\section{Linear response in the vector channel}

Having determined the effective vertices, we turn to evaluation of the
complete polarization tensor, which according to Eqs. (\ref{Pi}), (\ref{Qmu})
is given by%
\begin{equation}
\Pi_{\mu\nu}\left(  q\right)  =\Lambda_{\mu\nu}\left(  q\right)
-\frac{2\Delta}{q_{\lambda}\Lambda^{\lambda}\left(  q\right)  }\ \Lambda_{\mu
}\left(  q\right)  \Lambda_{\nu}\left(  q\right)  , \label{Pi1}%
\end{equation}
With the aid of relation (\ref{1}) this expression can be transformed to the
form which explicitly exhibits the continuity of the current:%
\[
\Pi_{\mu\nu}\left(  q\right)  =\Lambda_{\mu\nu}\left(  q\right)  -\frac
{1}{q^{\lambda}q^{\delta}\Lambda_{\lambda\delta}\left(  q\right)  }%
\ q^{\alpha}q^{\beta}\Lambda_{\mu\alpha}\left(  q\right)  \Lambda_{\beta\nu
}\left(  q\right)  .
\]
Indeed, this form obeys $\Pi_{\mu\nu}q^{\nu}=q^{\mu}\Pi_{\mu\nu}=0$. Using
this fact we decompose the polarization tensor (\ref{Pi1}) into the sum of
longitudinal (with respect to $\mathbf{q}$) and transverse components%
\begin{equation}
\Pi^{\mu\nu}\left(  q\right)  =\Pi_{l}\left(  q\right)  \left(  1,\frac
{\omega}{\mathsf{q}}\mathbf{n}\right)  ^{\mu}\left(  1,\frac{\omega
}{\mathsf{q}}\mathbf{n}\right)  ^{\nu}+\Pi_{t}\left(  q\right)  g^{\mu
i}\left(  \delta^{ij}-\mathsf{n}^{i}\mathsf{n}^{j}\right)  g^{j\nu}.
\label{mnLT}%
\end{equation}
Here $\mathbf{n}=\mathbf{q/}\mathsf{q}$ is a unit vector; the longitudinal and
transverse polarization functions are defined as%
\begin{equation}
\Pi_{l}\left(  q\right)  =\Pi^{00}\left(  q\right)  ,\ \ \ \ \Pi_{t}\left(
q\right)  =\frac{1}{2}\left(  \delta^{ij}-\mathsf{n}^{i}\mathsf{n}^{j}\right)
\Pi^{ij}\left(  q\right)  . \label{PiLT}%
\end{equation}

As it follows from the definition (\ref{MN}), $\Lambda_{\mu\nu}\left(
\omega,\mathsf{q}\right)  $ represents the polarization tensor in the
so-called one-loop approximation. The transverse external field does not
influence the anomalous self-energy of a quasiparticle \cite{AGD}, therefore
the transverse polarization function is given by the one-loop expression
(\ref{T}). The calculation yields the well known expression
\begin{gather}
~\Pi_{t}\left(  \omega,\mathsf{q}\right)  =-\frac{1}{4M^{\ast2}}\int
\frac{d^{3}\mathsf{p}}{\left(  2\pi\right)  ^{3}}~\mathsf{p}^{2}\sin^{2}%
\theta\left[  \left(  1-\frac{\xi_{\mathsf{p}}\xi_{\mathbf{p+q}}+\Delta^{2}%
}{\varepsilon_{\mathsf{p}}\varepsilon_{\mathbf{p+q}}}\right)  \Phi_{+}\right.
\nonumber\\
\left.  +\left(  ~1+\frac{\xi_{\mathsf{p}}\xi_{\mathbf{p+q}}+\Delta^{2}%
}{\varepsilon_{\mathsf{p}}\varepsilon_{\mathbf{p+q}}}\right)  \Phi_{-}\right]
. \label{T}%
\end{gather}
Compare this to the linear response of superconducting electrons to external
transverse electromagnetic field \cite{AGD}.

In contrast, the longitudinal external field varies the order parameter in the
system. This physically manifests itself as the collective acoustic excitation
in the condensate. Therefore the longitudinal polarization function has the
additional contribution, which takes into account the collective motion of the
condensate providing a continuity of the current in the system. We obtain%
\begin{equation}
\Pi_{l}\left(  q\right)  =\Lambda_{00}\left(  q\right)  -\frac{2\Delta
}{q_{\lambda}\Lambda^{\lambda}\left(  q\right)  }\ \Lambda_{0}\left(
q\right)  \Lambda_{0}\left(  q\right)  \label{PILE}%
\end{equation}
According to Eq.(\ref{qiLi}), up to accuracy $V_{F}^{3}$ this can be
simplified as%
\begin{equation}
\Pi_{l}\left(  \omega,\mathsf{q}\right)  \simeq\Lambda_{00}\left(
\omega,\mathsf{q}\right)  -\frac{2\Delta\omega}{\omega^{2}-\mathsf{q}^{2}%
c_{s}^{2}}\ \Lambda_{0}\left(  \omega,\mathsf{q}\right)  \label{PiL}%
\end{equation}
or, identically,
\begin{align*}
\Pi_{l}\left(  q\right)   &  =-\int\frac{d^{3}\mathsf{p}}{2\left(
2\pi\right)  ^{3}}\left[  \left(  1-\frac{\xi_{\mathsf{p}}\xi_{\mathbf{p+q}%
}-\Delta^{2}}{\varepsilon_{\mathsf{p}}\varepsilon_{\mathbf{p+q}}}\right)
\Phi_{+}-~\left(  1+\frac{\xi_{\mathsf{p}}\xi_{\mathbf{p+q}}-\Delta^{2}%
}{\varepsilon_{\mathsf{p}}\varepsilon_{\mathbf{p+q}}}\right)  \Phi_{-}\right]
\\
&  +\frac{2\Delta^{2}\omega^{2}}{\omega^{2}-\mathsf{q}^{2}c_{s}^{2}}\int
\frac{d^{3}\mathsf{p}}{2\left(  2\pi\right)  ^{3}}\ \frac{1}{\varepsilon
_{\mathsf{p}}\varepsilon_{\mathbf{p+q}}}\left(  \Phi_{+}+\Phi_{-}\right)  .
\end{align*}
By the use of the following trick%
\[
\frac{\omega^{2}}{\omega^{2}-\mathsf{q}^{2}c_{s}^{2}}\equiv1+\frac
{\mathsf{q}^{2}c_{s}^{2}}{\omega^{2}-\mathsf{q}^{2}c_{s}^{2}}%
\]
we can write%
\begin{align*}
\Pi_{l}\left(  q\right)   &  =-\int\frac{d^{3}\mathsf{p}}{2\left(
2\pi\right)  ^{3}}\left[  \left(  1-\frac{\xi_{\mathsf{p}}\xi_{\mathbf{p+q}%
}+\Delta^{2}}{\varepsilon_{\mathsf{p}}\varepsilon_{\mathbf{p+q}}}\right)
\Phi_{+}-~\left(  1+\frac{\xi_{\mathsf{p}}\xi_{\mathbf{p+q}}+\Delta^{2}%
}{\varepsilon_{\mathsf{p}}\varepsilon_{\mathbf{p+q}}}\right)  \Phi_{-}\right]
\\
&  +\frac{\mathsf{q}^{2}c_{s}^{2}}{\omega^{2}-\mathsf{q}^{2}c_{s}^{2}}%
\int\frac{d^{3}\mathsf{p}}{2\left(  2\pi\right)  ^{3}}\ \frac{2\Delta^{2}%
}{\varepsilon_{\mathsf{p}}\varepsilon_{\mathbf{p+q}}}\left(  \Phi_{+}+\Phi
_{-}\right)  .
\end{align*}
and substitute $\xi_{\mathbf{p+q}}$ as given by Eq. (\ref{b}). Then with the
aid of the series expansion of the integrand up to accuracy $\left(
\mathsf{q}V_{F}/\Delta\right)  ^{3}$, and the averaging over angles after some
simplifications we arrive at the following result:%

\begin{align}
\Pi_{l}\left(  q\right)   &  =-\frac{1}{2T}\frac{\mathsf{p}_{F}^{2}}{2\pi^{2}%
}\int~d\mathsf{p}\left(  1-\frac{1}{2}\frac{\omega}{\mathsf{q}u_{\mathsf{p}}%
}\ln\frac{\omega+qu_{\mathsf{p}}}{\omega-qu_{\mathsf{p}}}\right)  \left(
1-\tanh^{2}\frac{\varepsilon_{\mathbf{p}}}{2T}\right) \nonumber\\
&  +\frac{\mathsf{q}^{2}c_{s}^{2}}{\left(  \omega+i0\right)  ^{2}%
-\mathsf{q}^{2}c_{s}^{2}}\frac{\mathsf{p}_{F}^{2}}{2\pi^{2}}\int
~d\mathsf{p}\frac{\Delta^{2}}{\varepsilon_{\mathbf{p}}^{3}}\tanh
\frac{\varepsilon_{\mathbf{p}}}{2T}\nonumber\\
&  -\frac{\mathsf{q}^{2}c_{s}^{2}}{\left(  \omega+i0\right)  ^{2}%
-\mathsf{q}^{2}c_{s}^{2}}\frac{1}{2T}\frac{\mathsf{p}_{F}^{2}}{2\pi^{2}}%
\int~d\mathsf{p}\frac{\Delta^{2}}{\varepsilon_{\mathbf{p}}^{2}}~\nonumber\\
&  \times\left(  1-\frac{1}{2}\frac{\omega}{\mathsf{q}u_{\mathsf{p}}}\ln
\frac{\omega+qu_{\mathsf{p}}}{\omega-qu_{\mathsf{p}}}\right)  \left(
1-\tanh^{2}\frac{\varepsilon_{\mathbf{p}}}{2T}\right)  , \label{PIL}%
\end{align}
where%
\[
u_{\mathsf{p}}\equiv V_{F}\frac{\xi_{\mathsf{p}}}{\varepsilon_{\mathsf{p}}}.
\]
is the velocity of a quasiparticle.

As given by this expression, contributions into the longitudinal polarization
of a superfluid fermion system arise both due to the motion of quasiparticles
(the first line) and the collective motion of the condensate (the second
line). There is also the mixed term given in the last two lines of Eq.
(\ref{PIL}).

At $T>T_{c}$ the energy gap vanishes and Eq. (\ref{PIL}) reduces to the well
known expression for the longitudinal polarization of a normal Fermi gas:
\[
\Pi_{l}\left(  q;T>T_{c}\right)  =-\frac{p_{F}M^{\ast}}{\pi^{2}}\left(
1-\frac{1}{2}\frac{\omega}{\mathsf{q}V_{F}}\ln\frac{\omega+qV_{F}}%
{\omega-qV_{F}}\right)  .
\]

The quasiparticle contribution and the mixed term vanish at zero temperature,
when all the fermions are paired and no quasiparticle excitations exist.
Therefore, at zero temperature, the longitudinal polarization arises only due
to the collective motion of the condensate. We obtain%
\begin{equation}
\Pi_{l}\left(  q;T=0\right)  =-\frac{p_{F}M^{\ast}}{\pi^{2}}\frac
{\mathsf{q}^{2}c_{s}^{2}}{\left(  \omega+i0\right)  ^{2}-\mathsf{q}^{2}%
c_{s}^{2}}. \label{T0}%
\end{equation}

\subsection{Imaginary part}

The imaginary part of the polarization function deserves of a special
consideration because it contains a complete information on the energy losses
and scattering of external particles in the medium. Here we have to
distinguish the space-like, $\mathbf{q}^{2}>\omega^{2},$ and time-like,
$\mathbf{q}^{2}<\omega^{2},$ kinematical domains. One can easily see that the
polarization function (\ref{PIL}) has no imaginary part in the time-like
kinematical domain. This due to the fact that the polarization function, as
given by Eq. (\ref{PIL}), is calculated to accuracy $V_{F}^{2}$. As we shall
see, in the time-like kinematical domain, the imaginary part is proportional
to $V_{F}^{4}$.

\subsubsection{Space-like momentum transfer}

In the space-like kinematical domain, $\mathbf{q}^{2}>\omega^{2}$, the
longitudinal polarization function has the imaginary part due to the pole
contributions of the collective mode at $\omega^{2}=\mathsf{q}^{2}c_{s}^{2}$
and due to the Landau damping at $\omega<\mathsf{q}u_{\mathsf{p}}$. The
corresponding expressions can be readily found from Eq. (\ref{PIL}). At finite
temperature this expression has a cumbersome form and represents no practical
interest in the nonrelativistic case. Therefore we shall restrict our
consideration of the space-like momentum transfer to the case of zero
temperature. From Eq. (\ref{T0}) we readily find%
\[
\operatorname{Im}\Pi_{l}\left(  q;T=0\right)  =-\frac{p_{F}M^{\ast}}{\pi^{2}%
}\mathsf{q}^{2}c_{s}^{2}\delta\left(  \omega^{2}-\mathsf{q}^{2}c_{s}%
^{2}\right)  \operatorname*{sign}\omega
\]
Since no quasiparticles exist at zero temperature, the imaginary part arises
only due to excitations of the collective motion of the condensate. This form
satisfies the $f$-sum rule%

\[
\int_{0}^{\mathsf{\infty}}\omega\operatorname{Im}\Pi_{l}\left(  q;T=0\right)
d\omega=\frac{p_{F}^{3}}{6\pi M^{\ast}}\mathsf{q}^{2}%
\]

\subsubsection{Time-like momentum transfer}

We focus now on the imaginary part of the retarded polarization function at
the time-like momentum transfer, $\mathbf{q}^{2}<\omega^{2}$, important for
applications. As mentioned above, this imaginary part needs the calculation of
corrections of the order $V_{F}^{4}$.

At $\omega>2\Delta$ and $\mathsf{q}<\omega$ the imaginary part in Eqs.
(\ref{L00})-(\ref{Li}) and (\ref{T}) arises from the pole of the function
$\Phi_{+}$ at $\omega=\varepsilon_{\mathbf{p+q}}+\varepsilon_{\mathsf{p}}$.
Calculation of the imaginary part of the transverse polarization function
(\ref{T}) is trivial. We get
\begin{equation}
~\operatorname{Im}\Pi_{t}\left(  \omega,\mathsf{q}\right)  =\frac{1}{15\pi
}V_{F}^{4}p_{F}M^{\ast}\frac{\mathsf{q}^{2}\Delta^{2}\Theta\left(  \omega
^{2}-\mathsf{q}^{2}\right)  }{\omega^{3}\sqrt{\omega^{2}-4\Delta^{2}}}%
\tanh\frac{\omega}{4T}+O\left(  \allowbreak V_{F}^{4}\frac{\mathsf{q}^{2}%
}{\mathsf{p}_{F}^{2}},V_{F}^{5}\right)  \label{IT}%
\end{equation}
where $\Theta\left(  x\right)  $ is the Heaviside step function.

Calculation of the imaginary part of the longitudinal polarization function is
more complicated. First of all, we evaluate the one-loop integrals up to
accuracy $V_{F}^{4}$. \ This gives:
\begin{equation}
\Lambda_{00}(\omega,\mathbf{q})\simeq a_{0}(\omega,\mathbf{q})+a_{2}%
(\omega,\mathbf{q})V_{F}^{2}+O\left(  \allowbreak V_{F}^{3}\frac
{\mathsf{q}^{2}}{\mathsf{p}_{F}^{2}},V_{F}^{5}\right)  , \label{L00exp}%
\end{equation}%
\begin{equation}
\Lambda_{0}(\omega,\mathbf{q})\simeq b_{0}(\omega,\mathbf{q})+b_{2}%
(\omega,\mathbf{q})V_{F}^{2}+O\left(  \allowbreak V_{F}^{3}\frac
{\mathsf{q}^{2}}{\mathsf{p}_{F}^{2}},V_{F}^{5}\right)  , \label{L0exp}%
\end{equation}%
\begin{equation}
q_{i}\Lambda_{i}(\omega,\mathbf{q})\simeq c_{2}(\omega,\mathbf{q})V_{F}%
^{2}+O\left(  \allowbreak V_{F}^{3}\frac{\mathsf{q}^{2}}{\mathsf{p}_{F}^{2}%
},V_{F}^{5}\right)  , \label{Liexp}%
\end{equation}
where%
\begin{align}
\operatorname{Im}a_{0}  &  =-\frac{\mathsf{p}_{F}M^{\ast}}{\pi}\frac
{\Delta^{2}}{\omega\sqrt{\omega^{2}-4\Delta^{2}}}\tanh\frac{\omega}%
{4T},\ \nonumber\\
\operatorname{Im}a_{2}  &  =-\frac{\mathsf{p}_{F}M^{\ast}}{\pi}\frac
{\mathsf{q}^{2}}{3\omega^{2}}\frac{\Delta^{2}}{\omega\sqrt{\omega^{2}%
-4\Delta^{2}}}\tanh\frac{\omega}{4T}\nonumber\\
&  \times\left(  \frac{2\omega^{2}-2\Delta^{2}}{\omega^{2}-4\Delta^{2}}%
-\frac{\omega^{2}-4\Delta^{2}}{16T^{2}}\left(  1-\tanh^{2}\frac{1}{4T}%
\omega\right)  \right)  , \label{a}%
\end{align}%
\begin{align}
\operatorname{Re}b_{0}  &  =-\frac{\mathsf{p}_{F}M^{\ast}}{\pi^{2}}%
\mathcal{P}\int_{\Delta}^{\infty}d\varepsilon\frac{\Delta\omega}%
{\sqrt{\varepsilon^{2}-\Delta^{2}}}\frac{1}{4\varepsilon^{2}-\omega^{2}}%
\tanh\frac{\omega}{4T},\nonumber\\
\operatorname{Im}b_{0}  &  =-\frac{\mathsf{p}_{F}M^{\ast}}{2\pi}\frac{\Delta
}{\sqrt{\omega^{2}-4\Delta^{2}}}\tanh\frac{\omega}{4T},\nonumber\\
\operatorname{Im}b_{2}  &  =-\frac{\mathsf{p}_{F}M^{\ast}}{2\pi}%
\frac{\mathsf{q}^{2}}{3\omega^{2}}\frac{\Delta}{\sqrt{\omega^{2}-4\Delta^{2}}%
}\tanh\frac{\omega}{4T}\nonumber\\
&  \times\left(  \frac{2\Delta^{2}+\omega^{2}}{\omega^{2}-4\Delta^{2}%
}\allowbreak\allowbreak-\frac{\left(  \omega^{2}-4\Delta^{2}\right)  }%
{16T^{2}}\left(  1-\tanh^{2}\frac{\omega}{4T}\right)  \allowbreak\right)  ,
\label{bb}%
\end{align}%
\begin{align}
\operatorname{Re}c_{2}  &  =-\frac{2}{3}\mathsf{q}^{2}\frac{p_{F}M^{\ast}}%
{\pi^{2}}\mathcal{P}\int_{\Delta}^{\infty}d\varepsilon\frac{\Delta}%
{\sqrt{\varepsilon^{2}-\Delta^{2}}}\frac{1}{4\varepsilon^{2}-\omega^{2}}%
\tanh\frac{\omega}{4T},\nonumber\\
\operatorname{Im}c_{2}  &  =-\frac{\mathsf{p}_{F}M^{\ast}}{\pi}\frac
{\mathsf{q}^{2}}{3\omega}\frac{\Delta}{\sqrt{\omega^{2}-4\Delta^{2}}}%
\tanh\frac{\omega}{4T}.\ \label{c}%
\end{align}
In the above, symbol $\mathcal{P}$ means the principal value of the integral.

As already mentioned, at $\mathbf{q}^{2}<\omega^{2}$, the imaginary part of
$\Pi_{l}$ is small as compared to the contributions evaluated in previous
sections to accuracy $V_{F}^{3}$. Therefore, instead of Eq. (\ref{PiL}) we
start from the exact form of the longitudinal polarization function, as given
by Eq. (\ref{PILE}) in order to obtain the following series expansion over
powers of $V_{F}$%
\[
\Pi_{l}\simeq\allowbreak\left(  a_{0}-2\frac{\Delta}{\omega}b_{0}\right)
+\left(  a_{2}-2\frac{\Delta}{\omega}\left(  b_{2}+\frac{1}{\omega}%
c_{0}\right)  \right)  V_{F}^{2}-\allowbreak2\frac{\Delta}{\omega^{3}}%
\frac{c_{2}^{2}}{b_{0}}V_{F}^{4}...
\]
Thus we have%
\begin{align*}
\operatorname{Im}\Pi_{l}  &  \simeq\allowbreak\left(  \operatorname{Im}%
a_{0}-2\frac{\Delta}{\omega}\operatorname{Im}b_{0}\right) \\
&  +V_{F}^{2}\left(  \operatorname{Im}a_{2}-2\frac{\Delta}{\omega
}\operatorname{Im}b_{2}-2\frac{\Delta}{\omega^{2}}\operatorname{Im}%
c_{0}\right)  -\allowbreak V_{F}^{4}\frac{2\Delta}{\omega^{3}}%
\operatorname{Im}\left(  \frac{c_{2}}{b_{0}}c_{2}\right)  ...
\end{align*}
Upon substituting Eqs. (\ref{a}), (\ref{bb}), (\ref{c}) in the first and
second terms of this expression we find that these leading terms mutually
cancel
\[
\operatorname{Im}a_{0}-2\frac{\Delta}{\omega}\operatorname{Im}b_{0}=0,
\]%
\[
\operatorname{Im}a_{2}-2\frac{\Delta}{\omega}\operatorname{Im}b_{2}%
-2\frac{\Delta}{\omega^{2}}\operatorname{Im}c_{0}=0,
\]
in accordance with the results obtained in previous sections.

In the third term we find identically%

\[
\frac{c_{2}}{\omega b_{0}}\equiv\frac{2}{3}\frac{q^{2}}{\omega^{2}},
\]
and thus%
\[
\operatorname{Im}\Pi_{l}\simeq\allowbreak-\allowbreak4\frac{\Delta}{\omega
^{2}}\frac{V^{4}}{3}\frac{q^{2}}{\omega^{2}}\operatorname{Im}c_{2}%
\]
We finally obtain the following lowest-order expression%
\begin{equation}
\mathrm{Im}\Pi_{l}\left(  \omega,\mathbf{q}\right)  \simeq\frac{4}{\pi}%
c_{s}^{4}\mathsf{p}_{F}M^{\ast}\allowbreak\allowbreak\frac{\mathsf{q}%
^{4}\Delta^{2}\Theta\left(  \omega-2\Delta\right)  }{\omega^{5}\sqrt
{\omega^{2}-4\Delta^{2}}}\tanh\frac{\omega}{4T} \label{ImV}%
\end{equation}
valid at $\mathsf{q}<\omega$.

We see that, in comparison with the one-loop approximation, as given by Eqs.
(\ref{L00exp}) and (\ref{a}), the self-consistent result (\ref{ImV}) is
suppressed by many orders of magnitude. The suppression factor is $c_{s}^{4}$
$\sim10^{-4}$.

\section{Neutrino emission at the pair recombination}

As an application of the obtained results we consider the neutrino-pair
emission through neutral weak currents occurring at the recombination of
quasiparticles into the condensate. The process is kinematically allowed due
to the existence of a superfluid energy gap $\Delta$, which admits the
quasiparticle transitions with time-like momentum transfer $q=\left(
\omega,\mathbf{q}\right)  $, as required by the final neutrino pair.

We consider the total energy which is emitted into neutrino pairs per unit
volume and time which is given by the following formula (see details e.g. in
\cite{L01}):
\begin{equation}
Q=\frac{G_{F}^{2}}{8}\sum_{\nu}\int\;\frac{\omega}{\exp\left(  \frac{\omega
}{T}\right)  -1}2\mathrm{Im}\Pi_{\mathrm{weak}}^{\mu\nu}\left(  q\right)
\mathrm{Tr}\left(  l_{\mu}l_{\nu}^{\ast}\right)  \frac{d^{3}q_{1}}{2\omega
_{1}(2\pi)^{3}}\frac{d^{3}q_{2}}{2\omega_{2}(2\pi)^{3}}, \label{Q}%
\end{equation}
where $G_{F}$ is the Fermi coupling constant, $l_{\mu}$ is the neutrino weak
current, and $\Pi_{\mathrm{weak}}^{\mu\nu}$ is the retarded weak polarization
tensor of the medium. The integration goes over the phase volume of neutrinos
and antineutrinos of total energy $\omega=\omega_{1}+\omega_{2}$ and total
momentum $\mathbf{q=q}_{1}+\mathbf{q}_{2}$. The symbol $\sum_{\nu}%
$\ \ indicates that summation over the three neutrino types has to be performed.

We shall consider the $^{1}S_{0}$ pairing of neutrons, which takes place in
the superfluid crust of neutron stars. In this case, the total spin of a bound
pair is zero and the axial-vector weak interaction occurs only due to small
relativistic effects which are usually omitted in the calculations
\cite{FRS76}. Therefore we concentrate on the vector part of the weak
interactions. In the vector channel, the weak polarization tensor is given by
$\Pi_{\mathrm{weak}}^{\mu\nu}\left(  q\right)  =C_{V}^{2}\Pi^{\mu\nu}\left(
q\right)  $, where $C_{V}$ is the vector weak coupling constant of a neutron,
and $\Pi^{\mu\nu}\left(  q\right)  $ stands for the polarization tensor as
given by Eq. (\ref{mnLT}).

By inserting $\int d^{4}q\delta^{\left(  4\right)  }\left(  q-q_{1}%
-q_{2}\right)  =1$ in this equation, and making use of the Lenard's integral
\
\[
\int\frac{d^{3}k_{1}}{2\omega_{1}}\frac{d^{3}k_{2}}{2\omega_{2}}%
\delta^{\left(  4\right)  }\left(  q-q_{1}-q_{2}\right)  \mathrm{Tr}\left(
l^{\mu}l^{\nu\ast}\right)  =\frac{4\pi}{3}\left(  q_{\mu}q_{\nu}-q^{2}%
g_{\mu\nu}\right)  \Theta\left(  q^{2}\right)  \Theta\left(  \omega\right)  ,
\]
where $\Theta(x)$ is the Heaviside step function, and $g_{\mu\nu
}=\mathsf{\mathrm{diag}}(1,-1,-1,-1)$ is the signature tensor, we can write%
\[
Q=\frac{1}{48\pi^{4}}G_{F}^{2}C_{V}^{2}\mathcal{N}_{\nu}\int_{0}^{\infty
}d\omega\int_{0}^{\omega}d\mathsf{q}\;\mathsf{q}^{2}\frac{\omega}{\exp\left(
\frac{\omega}{T}\right)  -1}\mathrm{Im}\Pi^{\mu\nu}\left(  q\right)  \left(
q_{\mu}q_{\nu}-q^{2}g_{\mu\nu}\right)  ,
\]
where $\mathcal{N}_{\nu}=3$ is the number of neutrino flavors.

Due to the current conservation $q_{\mu}q_{\nu}\mathrm{Im}\Pi^{\mu\nu}\left(
q\right)  =0$, an we find%

\[
Q=\frac{C_{V}^{2}G_{F}^{2}}{48\pi^{4}}\mathcal{N}_{\nu}\int_{0}^{\infty
}d\omega\int_{0}^{\omega}d\mathsf{q}\;\mathsf{q}^{2}\;\frac{\left(  \omega
^{2}-\mathsf{q}^{2}\right)  \omega}{\exp\left(  \frac{\omega}{T}\right)
-1}\left[  \left(  \frac{\omega^{2}}{\mathsf{q}^{2}}-1\right)  \mathrm{Im}%
\Pi_{l}\left(  q\right)  +2\mathrm{Im}\Pi_{t}\left(  q\right)  \right]  ,
\]
where the functions $\mathrm{Im}\Pi_{l,t}\left(  q\right)  $ are given by Eqs.
(\ref{IT}) and (\ref{ImV}).

After simple calculations the neutrino energy losses are found to be:
\begin{equation}
Q=\frac{536\mathcal{N}_{\nu}}{42\,525\pi^{5}}V_{F}^{4}G_{F}^{2}C_{V}%
^{2}\mathsf{p}_{F}M^{\ast}T^{7}y^{2}\int_{0}^{\infty}\ \frac{z^{4}}{\left(
e^{z}+1\right)  ^{2}}dx \label{QQ}%
\end{equation}
where $y=\Delta/T$, $z=\sqrt{x^{2}+y^{2}}$.

We see that the neutrino radiation via the vector weak currents in the
nonrelativistic system is suppressed by several orders of magnitude with
respect to that predicted in \cite{FRS76}. The suppression factor is $\sim
V_{F}^{4}$. This result is consistent with the result obtained in Ref.
\cite{LP06}, by the use of the Fermi Golden rule.

\section{Summary and conclusion}

We have developed a unified approach for calculating the longitudinal response
function of superfluid fermion system at finite temperatures. The effective
vertex in the vector channel is analytically expressed in terms of the order
parameter, and other macroscopic parameters of the superfluid system, without
of using of an explicit form of the pairing interaction. The general
expression for the vertex function is given by Eq. (\ref{GAMef}). When
evaluated to accuracy $V_{F}^{2}\ll1$, this expression becomes very simple and
takes the form, as given by Eqs. (\ref{GAM0}), (\ref{GAMq}). The latter
formally reproduce the result obtained in Ref. \cite{Nambu} for the case of
zero temperature and small transferred energy and momentum. In contrast, our
result is valid at finite temperatures and for $\omega$ and $\mathbf{q}$,
satisfying only $\omega\ll\mu,\left\vert \mathbf{q}\right\vert \ll
\mathsf{p}_{F}$. As we found to this accuracy $V_{F}^{2}\ll1$ the temperature
dependence of the effective vector vertex is restricted to the gap function
$\Delta\left(  T\right)  $. This justifies the approach used in
Ref.~\cite{LP06} where this form of the effective vertex was used in the
calculation of the neutrino energy losses.

Special attention was paid to preserving the Ward identity for the vertex
function and the relations for the polarization tensor components, that are
implied by the conservation of the current. The transverse and longitudinal
polarization functions, as represented by Eqs. (\ref{T}), (\ref{PIL}), are
calculated up to accuracy $V_{F}^{2}$.

The imaginary part of the response functions, as given by (\ref{T}),
(\ref{PIL}), arises due to the pole contribution of the acoustic-like
collective mode and due to the Landau damping. Both contributions vanish at
the time-like momentum transfer, $\omega>2\Delta$ and $\mathsf{q}<\omega$. In
this kinematical domain, the imaginary part of the polarization functions is
due to breaking of the correlated pairs and recombination of the
quasiparticles back into the condensate. The corresponding analytical
expressions are found to be proportional to $V_{F}^{4}$, as given by Eqs.
(\ref{IT}), (\ref{ImV}).

As an application of obtained results we have calculated the neutrino-pair
emission through neutral weak currents occurring at the recombination of
quasiparticles into the condensate. We found the rate of neutrino energy
losses in the vector channel is suppressed as compared to the one-loop results
by a factor $V_{F}^{4}$. The magnitude of the suppression is in accordance
with the one predicted in Ref.~\cite{LP06}. Thus the neutrino emission from
the singlet-correlated neutron matter is mainly due to the axial-vector
currents. As is well known \cite{FRS76}, the corresponding neutrino energy
losses are proportional to $V_{F}^{2}$ and thus also small.  Apparently, the
modifications to the neutrino emission rate through the pair recombination
process, as found above, call for a detail reassessment of their role in the
late-time cooling of neutron stars.

\end{document}